\documentclass[epj]{svjour}
\usepackage{graphicx}
\usepackage{amsmath}
\usepackage{amssymb}
\usepackage[utf8]{inputenc} 
\usepackage{url} 
\usepackage{color}
\DeclareMathOperator{\arccot}{arccot}
\begin{document}
\title{The role of spatial curvature in constraining the Universe anisotropies across a Big Bounce}
\author{Eleonora Giovannetti\inst{1}\thanks {eleonora.giovannetti@uniroma1.it} and Giovanni Montani\inst{2,1}\thanks {giovanni.montani@enea.it}
}                     
%
%
\institute{Department of Physics, “Sapienza” University of Rome, P.le Aldo Moro, 5 (00185) Roma, Italy \and ENEA, Fusion and Nuclear Safety Department, C.R. Frascati, Via E. Fermi, 45 (00044) Frascati (RM), Italy}
%
%
\abstract{We study the implementation of Polymer Quantum Mechanics (PQM) to a system decomposed into a quasi-classical background and a small quantum subsystem, according to the original Vilenkin proposal. We develop the whole formalism in the momentum representation that is the only viable in the continuum limit of the polymer paradigm and we generalize the fundamental equations of the original Vilenkin analysis in the considered context. Then, we provide a Minisuperspace application of the theory, first considering a Bianchi I cosmology and then extending the analysis to a Bianchi IX model in the limit of small anisotropies. In both these cases, the quasi-classical background is identified with an isotropic bouncing Universe whereas the small quantum subsystem contains the anisotropic degrees of freedom. When the Big Bounce scenario is considered, we obtain that in the Bianchi I model the anisotropies standard deviation is regular at $t=0$ but still increases indefinitely, whereas in the presence of the harmonic Bianchi IX potential such same quantity is bounded and oscillate around a constant value. As a consequence, we demonstrate that the picture of a semiclassical isotropic Bounce can be extended to more general cosmological settings if the spatial curvature becomes relevant when the anisotropic degrees of freedom are still small quantum variables.}
%
\authorrunning{E. Giovannetti and G. Montani}
\titlerunning{Role of spatial curvature in constraining the Universe anisotropies across a Bounce}
\maketitle
\section{Introduction}
The application of the canonical quantization of gravity to the cosmological problem in the metric approach \cite{DeWitt,Kuchar81,Thiemann,CQG} has not implied the hoped-for removal of the initial singularity. In fact, the Wheeler-DeWitt equation is associated to quasi-classical states which follow the Einsteinian classical behaviour of the corresponding cosmological solutions \cite{Misner69,Isham75,Benini2006,PC}. A different situation came out from the cosmological application of Loop Quantum Gravity \cite{Rovelli,Thiemann}, especially for what concerns the isotropic Universe dynamics \cite{Ashtekar05,Ashtekar05I,Ashtekar06,Ashtekar11}. Actually, the quantum dynamics of an isotropic Universe is associated to quasi-classical states which outline a clear bouncing picture, i.e. the symmetrical reconnection of the collapsing and expanding branches of the dynamics in correspondence of a minimum Universe volume. A clear picture for a bouncing cosmology also emerges for more general homogeneous models, as discussed in \cite{BI,BII,BIX} (more in general, for bouncing models obtained in modified gravity theories, see for example \cite{F1,F2,F3}). However, this important feature of replacing the initial Big Bang with a primordial Big Bounce is affected by some criticisms \cite{C1,C2,Bojowald_2020} (for a more general critique point of view on the Loop Quantum Gravity approach see \cite{Nicolai}). A simpler approach is provided by the implementation of PQM \cite{Corichi} to the Minisuperspace \cite{Barca21}. This formulation is still able to induce a bouncing cosmology, both in a metric and in a connection approach. For a discussion on the emergence of the polymer quantum dynamics from the Loop Quantum Cosmology paradigm see \cite{Ashtekar01}. Also, see \cite{E} and \cite{Antonini_2019,Giovannetti_2019,Mantero18,Giovannetti_2022,Giovannetti_2022b} for many interesting cosmological scenarios in effective Loop Quantum/Polymer Cosmology respectively. 

Here we face a central theme about the real nature of the bouncing isotropic cosmology, namely the question concerning the quantum behaviour of the anisotropic degrees of freedom across the Big Bounce. Actually, the tendency of the Universe anisotropies to highly increase towards the primordial Bounce could affect the robustness of the picture proposed by the Ashtekar School. Some attempts to overcome this problem can be found in \cite{ani1,ani2,ani3} that led to the formulation of the so-called Ekpyrotic Cosmology \cite{ekp}. Our proposal of studying the anisotropies behaviour of a quantum Universe refers to the original Vilenkin idea, presented in \cite{Vilenkin89} (see also \cite{Maniccia22} for a review on this theme), in which the Minisuperspace is separated into a quasi-classical background and a small quantum subsystem (for a precise derivation about the physical meaning of the word ``small'' see \cite{Agostini17}).  The important difference between the present study and previous cosmological implementations of the Vilenkin idea \cite{Battisti09,DeAngelis20,Chiovoloni20,Montani21} consists of the semiclassical polymer dynamics of the background, which allows the emergence of a bouncing dynamics (pioneering works in the same spirit are \cite{Montani2018}). As a first step, we derive the proposed formulation in its general setting with the only assumption of dealing with a Minisuperspace scenario in the polymer formulation. In this respect, we generalize the equations obtained in \cite{Vilenkin89} by addressing the momentum representation, the only viable in the construction of a continuum limit of the theory \cite{Corichi}. This part of the analysis is rather challenging from a mathematical point of view and actually we perform it with the ansatz that all the momentum functions are series expandable. The second part of the paper is dedicated to the implementation of the derived formalism to the Bianchi I and Bianchi IX models, when the anisotropic degrees of freedom are regarded as small variables. Actually, we deal with a semiclassical bouncing isotropic dynamics on which small anisotropies live (freely in the case of the Bianchi I model and subjected to a harmonic potential for the Bianchi IX one). For the Bianchi I model, the Schr\"{o}dinger equation describing the anisotropic variables dynamics resembles that one of a free two-dimensional particle. However, the bouncing dynamics enters in the definition of the time variable and regularizes the divergences of the anisotropies towards the singularity. Despite this, the anisotropies mean value is not confined and the standard deviation monotonically increases along the dynamics. Analogous results are obtained even when the anisotropies are considered as discrete in the polymer formulation. Therefore, in this picture we can conclude that the anisotropies are not under control during the Universe collapse, i.e. the semiclassical isotropic bouncing cosmology loses its reliability. Furthermore, when the Universe anisotropies increase enough the Vilenkin idea can be no longer applied. 

In the second case, when we analyze a Bianchi IX model with small enough anisotropies the Schr\"{o}dinger equation acquires a harmonic potential term (due to the model spatial curvature). Actually, in the polymer representation we deal with a time-dependent pendulum. In this respect, according to \cite{Vilenkin89,Agostini17} the anisotropic variables phase space is small and so we can approximate the polymer-like kinetic term with the ordinary quadratic one, so that we get a viable time-dependent harmonic oscillator. Then, we construct a complete base of eigenfunctions for the system and we show that in such states (including the low-energy one) the average value of the anisotropies is exactly zero with a standard deviation that oscillates with a constant amplitude. Differently from the previous case, here we see that the anisotropic degrees of freedom remain limited in amplitude during the bouncing dynamics of the background. This analysis shows that the presence of a small spatial curvature in the Bianchi IX model at a quantum level guarantees the physical robustness of a semiclassical isotropic bouncing cosmology. In other words, if we start with a quasi-isotropic Universe in the collapsing branch, then the quantum spatial curvature is able to preserve the smallness of the anisotropies across the Bounce. Since we have demonstrated the validity of this statement in the case of a Bianchi IX Universe (that reduces to a closed Robertson-Walker dynamics in the isotropic limit), this behaviour could concern more general cosmological scenarios in the context of the so-called BKL conjecture \cite{BKL82}. 

The paper is structured as follows. In Sec. \ref{pol} we present PQM and in Sec. \ref{vil} we describe our original version of the Vilenkin approach in the polymer formalism. In Sec. \ref{VBI} we apply the polymer formalism to the Bianchi IX model and we solve the quasi-classical dynamics. Then, in Sec. \ref{BI} and Sec. \ref{BIX} we perform the quantum analysis of the Bianchi I and IX models respectively. Finally, in Sec. \ref{C} we report the main results and we present some concluding remarks. 

\section{The polymer representation of quantum mechanics}\label{pol}
PQM is an alternative representation of quantum mechanics introduced by Corichi in \cite{Corichi}. Its formalism is based on the assumption that the configurational variables are discrete. Hence, it results non-equivalent to the Schr\"{o}dinger representation and its main applications regard the investigation of cut-off physics effects in quantum gravity and primordial cosmology theories. 

\subsection{Polymer kinematics}

PQM can be introduced without making any reference to the standard Schr\"{o}dinger representation. By considering abstract kets $|\mu\rangle$ labelled by the real parameter $\mu\in\mathbb{R}$, a generic state in the Hilbert space $\mathcal{H}_{poly}$ can be defined through a finite linear combination of them, i.e.
\begin{equation}
|\psi\rangle=\sum_{i=1}^Na_i|\mu_i\rangle\,,
\end{equation}
where $\mu_i\in\mathbb{R},\;i=1,\dots,N\in\mathbb{N}$. The inner product can be chosen as
\begin{equation}
\langle\mu|\nu\rangle=\delta_{\mu\nu}\,,
\end{equation}
in order to guarantee the orthonormality between the basis kets. It can be demonstrated that such a Hilbert space $\mathcal{H}_{poly}$ is non-separable. 

The two fundamental operators on $\mathcal{H}_{poly}$ are the symmetric operator $\hat{\epsilon}$ that labels the kets and the unitary operator $\hat{s}(\lambda)$ with $\lambda\in\mathbb{R}$ that shifts them. Their action is  
\begin{equation}
\hat{\epsilon}|\mu\rangle:=\mu|\mu\rangle
\end{equation}
and
\begin{equation}
\hat{s}(\lambda)|\mu\rangle:=|\mu+\lambda\rangle\,,
\end{equation}
respectively. Since the kets $|\mu\rangle$ and $|\mu+\lambda\rangle$ are orthogonal $\forall\lambda$, the shift operator $\hat{s}(\lambda)$ is discontinuous in $\lambda$ and therefore no Hermitian operator can represent it by exponentiation. 

To give an explicit representation of these two main operators, let us consider a one-dimensional system $(q,p)$ in which the configurational coordinate $q$ has a discrete character. It is easy to see that in the $p$-polarization the shift operator acts as
\begin{equation}
\hat{s}(\lambda)\cdot\psi_{\mu}(p)= e^{\frac{i\lambda p}{\hslash}}e^{\frac{i\mu p}{\hslash}}=e^{\frac{i(\mu+\lambda) p}{\hslash}}=\psi_{\mu+\lambda}(p)\,,
\end{equation}
so $\hat{s}(\lambda)$ can be identified with the operator $e^{\frac{i\lambda\hat{p}}{\hslash}}$ but $\hat{p}$ cannot be defined rigorously. On the other hand, the operator $\hat{q}$ acts as a differential operator
\begin{equation}
\hat{q}\cdot\psi_{\mu}(p)= -i\frac{\partial}{\partial p} \psi_{\mu}(p)= \mu \psi_{\mu}(p)
\end{equation}
and corresponds to the label operator $\hat{\epsilon}$. We notice that the eigenvalues of $\hat{q}$ can be considered as a discrete set, since they label kets that are orthonormal $\forall\lambda$.  

\subsection{Polymer dynamics}{\label{Polymer}}

Let us consider a one-dimensional system described by the Hamiltonian
\begin{equation}
H=\frac{p^2}{2m}+V(q)
\end{equation}
in the $p$-polarization. In order to deal with a well-defined dynamics, we have to find a proper definition for the physical operators $\hat{p}$ and $\hat{q}$. The main point of the kinematical analysis performed above is that we have to find an approximate representation for $\hat{p}$ when $\hat{q}$ has a discrete character. The standard procedure consists in the introduction of a lattice with a constant spacing $\mu$
\begin{equation}
\gamma_{\mu}=\{q \in \mathbb{R} : q=n\mu, \; \forall\;  n \in \mathbb{Z}\}\,.
\end{equation}
In order to remain in the lattice, the permitted states $|\psi\rangle=\sum_{n}b_n|\mu_n\rangle\in\mathcal{H}_{\gamma_{\mu}}$ are such that $\mu_n=n\mu$. The action of the operator $e^{\frac{i\lambda\hat{p}}{\hslash}}$, after been restricted to the lattice, is well-defined and can be used to define an approximate version of $\hat{p}$
\begin{eqnarray}
\nonumber\hat{p}_{\mu}|\mu_n\rangle&&=\frac{\hslash}{2i\mu}[e^{\frac{i\mu\hat{p}}{\hslash}}-e^{-\frac{i\mu\hat{p}}{\hslash}}]|\mu_n\rangle=\\&&=\frac{\hslash}{2i\mu}(|\mu_{n+1}\rangle-|\mu_{n-1}\rangle)\,.
\end{eqnarray}
Actually, for $\mu p\ll \hslash$ one gets $p\backsim\sin(\mu p)/\mu=\hslash(e^{\frac{i\mu p}{\hslash}}-e^{-\frac{i\mu p}{\hslash}})/2i\mu$. Accordingly, for $\hat{p}^2$ we obtain
\begin{eqnarray}
\nonumber\hat{p}^2_{\mu}|\mu_n\rangle&&\equiv\hat{p}_{\mu}\cdot\hat{p}_{\mu}|\mu_n\rangle=\\\nonumber&&=\frac{\hslash^2}{4\mu^2}[-|\mu_{n-2}\rangle+2|\mu_{n}\rangle-|\mu_{n+2}\rangle]=\\&&=\frac{\hslash^2}{\mu^2}\sin^2(\mu p)|\mu_n\rangle\,.
\end{eqnarray}
We remind that $\hat{q}$ is well-defined, so the regularized version of the Hamiltonian is
\begin{equation}
\hat{H}_{\mu}:=\frac{1}{2m}\hat{p}_{\mu}^2+\hat{V}(q)
\end{equation}
and represents a symmetric and well-defined operator on $\mathcal{H}_{\gamma_{\mu}}$.

\section{Vilenkin approach in the polymer formalism}\label{vil}
In this section we present the procedure we developed to reproduce the Vilenkin picture in the polymer framework. The original Vilenkin proposal was firstly presented in \cite{Vilenkin89} and is about the probabilistic interpretation of the Universe wave function. In the same spirit, we start by separating the total Hamiltonian of the considered system in its classical and quantum parts 
\begin{equation}
\label{totH}
    H=H_0+H_q\,,H_0=p_ip_jH^{ij}(i\hslash\partial p)+V(i\hslash\partial p)\,.
\end{equation}
The basic idea is that the concept of time, and hence a probabilistic interpretation of the wave function, can be introduced only in a system with small quantum fluctuations. So, the fundamental requirement is dealing with some quasi-classical degrees of freedom and other quantum ones. In \eqref{totH}, $H_0$ is the quasi-classical part of the total Hamiltonian $H$ and $H_q$ the quantum one, whereas $H_{ij}$ is the Supermetric. In the case of one-dimensional systems we get 
\begin{equation}
\label{ham}
H=H_0+H_q\,,H_0=p^2g(i\hslash\partial p)+V(i\hslash\partial p)\,,
\end{equation}
We note that the use of the momentum representation is required to implement the polymer representation. Moreover, $p$ is multiplicative and the chosen ordering is the normal one. However, $H_0$ must satisfy some regularity criteria in order the formalism to be developed. In particular, it is required that
\begin{equation}
\label{alfa}
    g(i\hslash\partial p)=\sum_{n=0}^{+\infty}a_n(i\hslash)^n(\partial p)^n
\end{equation}
and
\begin{equation}
    V(i\hslash\partial p)=\sum_{n=0}^{+\infty}b_n(i\hslash)^n(\partial p)^n\,.
\end{equation}
Actually, every analytic function can be expanded in power series and hence the following procedure will be valid inside the radius of convergence.

By following the Vilenkin approach as in \cite{Vilenkin89}, we consider the Universe wave function to be the product of a quantum contribution $\chi(q,p)$ times a quasi-classical WKB one $\psi(p)$, i.e.
\begin{equation}    \Psi(p,q)=\psi(p)\chi(p,q)=A_k(p)e^{-i/\hslash S_k(p)}\chi(p,q)\,,
\end{equation}
in which we consider $p$ as a quasi-classical variable and $q$ as a quantum one. In order to guarantee that the quantum effects on the quasi-classical system are negligible, we impose 
\begin{equation}
    \frac{H_q\psi}{H_0\psi}=o(\hslash)\,,
\end{equation}
i.e. the quantum degrees of freedom constitute a small subsystem compared to the quasi-classical phase space. In addition, we make a Born-Oppenheimer hypothesis by considering the dependence of $\chi(p,q)$ from $p$ as parametric. Thanks to these hypotheses, we can derive the Vilenkin equations in the momentum space and demonstrate that the quasi-classical dynamics is completely described by a Hamilton-Jacobi together with a continuity equation and that a Schr\"{o}dinger equation emerges at a quantum level. 

We remark that in this scheme the semiclassical polymer substitution $p\rightarrow\sin{(\mu p)}/\mu$ can be easily implemented since the variable $p$ acts multiplicatively on the left in \eqref{totH}. Hence, the expression of the polymer-modified Hamiltonian becomes
\begin{equation}
    \label{hampol}
    H^{pol}=H_0^{pol}+H_q\,,H_0^{pol}=\frac{1}{\mu^2}\sin^2(\mu p)g(i\hslash\partial p)+V(i\hslash\partial p)\,.
\end{equation}

The first polymer Vilenkin equation represents the annihilation of the quasi-classical Hamiltonian $H_0^{pol}$ on the quasi-classical part of the Universe wave function $\psi$, i.e.
\begin{equation}
\label{Hzero}
H_0^{pol}\psi(p)=H_0^{pol}A(p)e^{-i/\hslash S(p)}=0\,.
\end{equation}
By using \eqref{alfa}, at the lowest order in $\hslash$ we have
\begin{align}
\nonumber
    \frac{\sin^2(\mu p)}{\mu^2}\sum_na_n\bigg(\frac{\partial S(p)}{\partial p}\bigg)^nA(p)e^{-i/\hslash S(p)}+\\+V(i\hslash\partial p)A(p)e^{-i/\hslash S(p)}=0
\end{align}
and after summing the series we get
\begin{equation}
\label{jacobi}
H_0^{pol}\bigg(p,\frac{\partial S}{\partial p}\bigg)= \frac{\sin^2(\mu p)}{\mu^2}g\bigg(\frac{\partial S}{\partial p}\bigg)+V(i\hslash\partial p)=0\,,
\end{equation}
i.e. the Hamilton-Jacobi equation. At the next order we obtain
\begin{equation}
\frac{\sin^2(\mu p)}{\mu^2}\sum_na_ni\hslash\frac{\partial}{\partial p}\bigg[n\,A(p)\bigg(\frac{\partial S(p)}{\partial p}\bigg)^{n-1}\bigg]e^{-i/\hslash S(p)}=0
\end{equation}
and by using \eqref{jacobi} we get
\begin{equation}
\label{continuity}
    \sum_na_n\frac{\partial}{\partial p}\bigg[n\,A(p)\bigg(\frac{\partial S(p)}{\partial p}\bigg)^{n-1}\bigg]=0
\end{equation}
that corresponds to the continuity equation. By solving both \eqref{jacobi} and \eqref{continuity} we can derive the quasi-classical action $S(p)$ and the amplitude $A(p)$ and then characterize the quasi-classical contribution to the Universe wave function $\psi(p)$ in the polymer scheme.

The second polymer Vilenkin equation leads to the emergence of a Schr\"{o}dinger dynamics at the first order in $\hslash$ and hence to the definition of a time variable for the system. In particular, it represents the annihilation of the total Hamiltonian $H^{pol}$ on the Universe wave function $\Psi(p,q)$, i.e.
\begin{equation}
\label{Htot}
H^{pol}\psi(p)\chi(p,q)=(H_0^{pol}+H_q)A(p)e^{-i/\hslash S(p)}\chi(p,q)=0\,.
\end{equation}
Thanks to the Born-Oppenheimer assumption we neglect the action of $H_q$ on $p$, so by using \eqref{alfa} we can write
\begin{align}
    \nonumber
     \frac{\sin^2(\mu p)}{\mu^2}\sum_na_n(i\hslash)^n(\partial p)^n[A(p)e^{-i/\hslash S(p)}\chi(q,p)]=\\=-A(p)e^{-i/\hslash S(p)}H_q\chi(q,p)
\end{align}
and at the first order in $\hslash$ we obtain
\begin{equation}
\label{2V}
     \frac{\sin^2(\mu p)}{\mu^2}\sum_na_ni\hslash \,n\bigg(\frac{\partial S}{\partial p}\bigg)^{n-1}\frac{\partial\chi}{\partial p}=-H_q\chi(q,p)\,\,
\end{equation}
in which we have used \eqref{jacobi} and \eqref{continuity}. Now, we can rewrite the l.h.s of \eqref{2V} as
\begin{align}
\nonumber
     \frac{\sin^2(\mu p)}{\mu^2}\sum_na_ni\hslash 
 \,n\frac{\partial\chi}{\partial p}\bigg(\frac{\partial S}{\partial p}\bigg)^{n-1}&=\\=\frac{\partial}{\frac{\partial S}{\partial p}}   \sum_na_ni\hslash\frac{\partial\chi}{\partial p}\bigg(\frac{\partial S}{\partial p}\bigg)^n&=i\hslash\frac{\partial\chi}{\partial p}\frac{\partial}{\frac{\partial S}{\partial p}} H_0^{pol}\bigg(\frac{\partial S}{\partial p}\bigg)\,,
\end{align}
thus obtaining
\begin{equation}
    i\hslash\frac{\partial\chi}{\partial p}\frac{\partial}{\frac{\partial S}{\partial p}}H_0^{pol}\bigg(\frac{\partial S}{\partial p}\bigg)=-H_q\chi\,.
\end{equation}
We also remind that $q=\partial S/\partial p$, so by definition we have
\begin{equation}
    \frac{\partial H_0^{pol}}{\frac{\partial S}{\partial p}}=\frac{\partial H_0^{pol}}{\partial q}=-\dot{p}/N
\end{equation}
where $N$ is the lapse function, and therefore we can rewrite \eqref{2V} as
\begin{equation}
    i\hslash\frac{\partial\chi}{\partial p}\frac{\partial p}{\partial t}=NH_q\chi\,.
\end{equation}
Hence, we have recovered a Schr\"{o}dinger equation for the system, i.e.
\begin{equation}
    i\hslash\frac{\partial\chi}{\partial t}=NH_q\chi\,.
\end{equation}

\section{Formulation for a Bianchi IX model}\label{VBI}
Our aim is to investigate the quantum behaviour of the anisotropies when PQM is implemented to a quasi-classical cosmological setting and hence a semiclassical Big Bounce emerges. In order to describe the primordial Universe as accurately as possible, let us consider a Bianchi IX model with a free scalar field $\phi$. In this respect, we consider the regime of small anisotropies far from the Bounce, in order to describe them as the quantum subsystem in the Vilenkin approach.

Let us start with the Bianchi IX Hamiltonian in the semiclassical polymer formulation
\begin{align}
\nonumber    H_{IX}^{pol}&=\frac{A^{-3/2}\sqrt{2}}{2}\bigg[-\frac{2}{3\sqrt{2}(4\pi)^2}\frac{A^2\sin^2(\mu p_A)}{\mu^2}+p_\phi^2+\\&+p_+^2+p_-^2+A^2(\beta_+^2+\beta_-^2)\bigg]\,,
    \label{HIX}
\end{align}
where $\beta_+,\beta_-$ are the anisotropies and $A=e^{2\alpha}$ ($\alpha$ being the isotropic Misner variable) is the degree of freedom related to the Universe area and chosen as discrete in the polymer framework. We remark that in \eqref{HIX} the Bianchi IX potential has been expanded up to the second order in the anisotropies \cite{gravitation} thanks to the Vilenkin hypothesis $\beta_+,\beta_-\ll1$. We also note that $8\pi G=1$ and some irrelevant constants have been taken out by canonically redefining the coordinates.

\begin{figure}[ht]
    \centering
\includegraphics[width=0.9\linewidth]{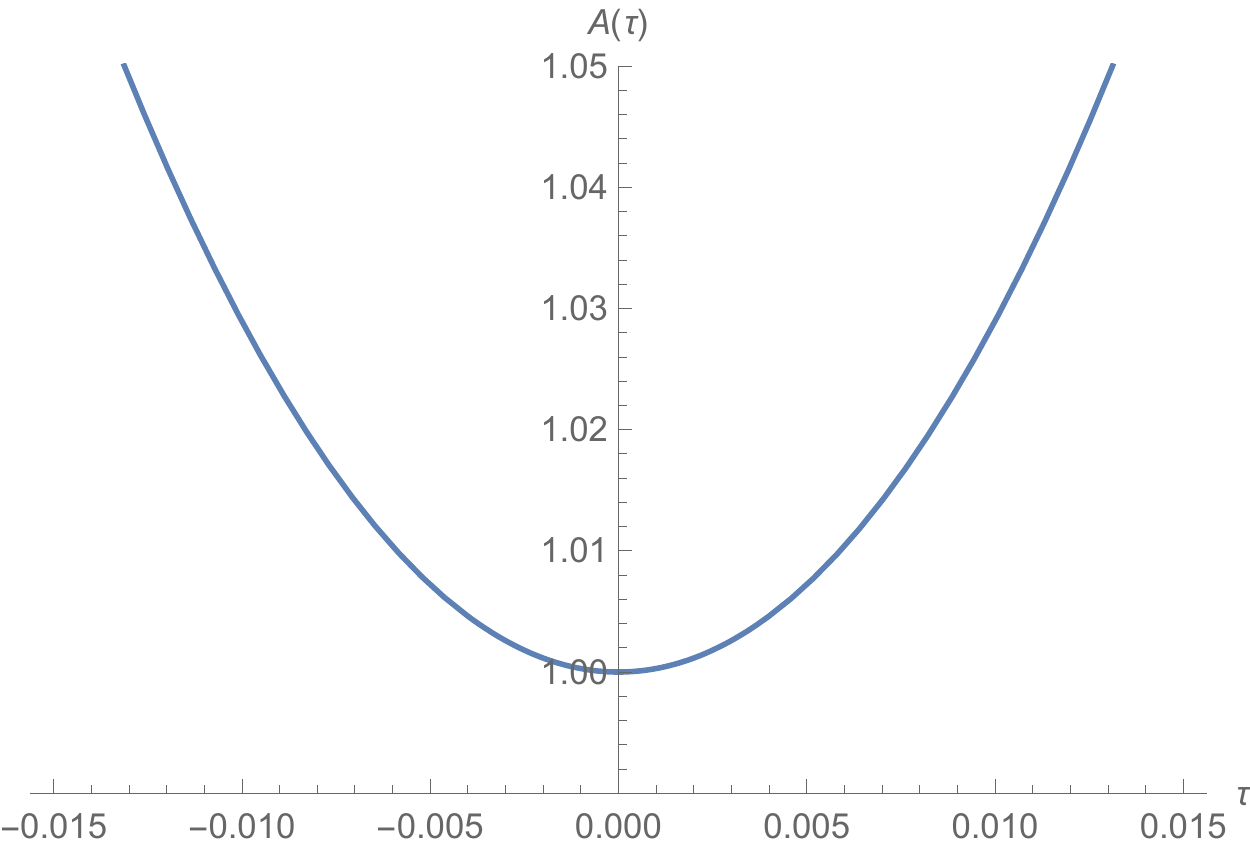}
    \caption{Plot of the bouncing solution $A(\tau)$ for $\mu=1/3$.}
    \label{bounce}
\end{figure}

By following the Vilenkin approach of Sec. \ref{vil}, we consider $A,\phi$ as the classical degrees of freedom and $\beta_+,\beta_-$ as the quantum ones. Therefore, the classical evolution will be dictated by
\begin{equation}
    H_0^{pol}=-\frac{2}{3\sqrt{2}(4\pi)^2}\frac{A^2\sin^2(\mu p_A)}{\mu^2}+p_\phi^2
\end{equation}
and the quantum evolution by
\begin{equation}
    H_q^{pol}=p_+^2+p_-^2+A^2(\beta_+^2+\beta_-^2)\,.
\end{equation}
The classical dynamics can be obtained by integrating the Hamilton equations
\begin{equation}
\begin{cases}
\displaystyle
\dot{A}=N\frac{\partial{H_0^{pol}}}{\partial p_A}=-\frac{1}{\sqrt{288}\pi^2}\frac{A^2}{\mu}\sin{(\mu p_A)}\cos{(\mu p_A)}
\\
\displaystyle
\dot{p_A}=-N\frac{\partial H_0^{pol}}{\partial A}=\frac{1}{\sqrt{288}\pi^2}\frac{A\sin^2(\mu p_A)}{\mu^2}
\end{cases}
\end{equation}
in which the time gauge has been fixed by choosing $N=2A^{3/2}/\sqrt{2}$. The analytical solutions for $(A,P_A)$ are
\begin{equation}
\label{sol}
\begin{cases}
\displaystyle
A(\tau)=\cosh{\bigg(\frac{\sin(\frac{3\mu\pi}{2})\tau}{\sqrt{288}\pi^2\mu}\bigg)}-\cos{\bigg(\frac{3\mu\pi}{2}\bigg)}\sinh\bigg(\frac{\sin(\frac{3\mu\pi}{2})\tau}{\sqrt{288}\pi^2\mu}\bigg)
\\
\displaystyle
p_A(\tau)=\frac{2}{\mu}\arccot\bigg[\exp{\bigg(-\frac{\sin(\frac{3\mu\pi}{2})\tau}{\sqrt{288}\pi^2\mu}\bigg)}\cot\bigg(\frac{3\mu\pi}{4}\bigg)\bigg]
\end{cases}
\end{equation}
with initial conditions $A(0)=1,\,p_A(0)=3\pi/2$. As expected, the Universe evolution traces a bouncing one, as shown in Fig. \ref{bounce}. We notice that this result would not have been obtained by simply polymerizing the isotropic Misner variable $\alpha$ \cite{Crin__2018,Giovannetti_2019}.

Now, let us derive the Vilenkin equations for the considered model. As discussed in Sec. \ref{vil}, regarding the first Vilenkin equation \eqref{Hzero} at the zero order we recover the Hamilton-Jacobi equation 
\begin{equation}
\label{jacobiIX}
   -\frac{2}{3\sqrt{2}(4\pi)^2}\frac{\sin^2(\mu p_A)}{\mu^2}\bigg(\frac{\partial S}{\partial p_A}\bigg)^2+p_\phi^2=0\,,
\end{equation}
whereas at the first order we have
\begin{equation}
\label{continuityIX}
    i\hslash\frac{\partial}{\partial p_A}\bigg(A\frac{\partial S}{\partial p_A}\bigg)=0\,,
\end{equation}
i.e. the continuity equation. Then, by solving \eqref{jacobiIX} and \eqref{continuityIX} we get the analytical expressions of the action $S$ and the amplitude $A$, i.e.
\begin{equation}
\begin{cases}
\displaystyle
S(p_A,p_\phi)=\pm\ln{(\tan(\mu p_A/2)^{2\pi\mu p_\phi\sqrt{6\sqrt{2}}})}
\\
\displaystyle
A(p_A,p_\phi)=\frac{\sin(\mu p_A)}{2\pi\mu p_\phi\sqrt{6\sqrt{2}}}
\end{cases}
\end{equation}
Hence, for the portion of the phase space in which $\partial S/\partial p_A>0$ the classical part of the Universe wave function results to be
\begin{equation}
\psi(p_A,p_\phi)=\frac{\sin(\mu p_A)}{2\pi\mu p_\phi\sqrt{6\sqrt{2}}}\exp{\bigg[-\frac{i}{\hslash} \ln{\bigg(\tan\bigg(\frac{\mu p_A}{2}\bigg)^{2\pi\mu p_\phi\sqrt{6\sqrt{2}}}\bigg)}\bigg]}
\end{equation}
in which it is encoded all the information on the quasi-classical phase space dynamics.

In order to obtain the full description of the Universe, we have to solve the second Vilenkin equation \eqref{Htot}. In particular, at the first order it emerges a Schr\"{o}dinger evolution 
\begin{equation}
    \label{S}
    \bigg[p_+^2+p_-^2-A^2(\tau)\hslash^2\bigg(\frac{\partial^2}{\partial p_+^2}+\frac{\partial^2}{\partial p_-^2}\bigg)\bigg]\chi(p_\pm,\tau)=i\hslash\frac{\partial \chi(p_\pm,\tau)}{\partial \tau}
\end{equation}
that resembles a time-dependent quantum harmonic oscillator in which the time parameter is defined through the relation
\begin{equation}
\label{tau}
    \frac{\partial}{\partial\tau}=N\frac{\partial}{\partial t}=\frac{2A^{3/2}}{\sqrt{2}}\frac{\partial}{\partial t}\,.
\end{equation}
To be accurate, the anisotropies are the real quantum degrees of freedom and therefore should be affected by the polymer modifications. In that case, \eqref{S} would be
\begin{align}
    \label{P}
   &\bigg[\frac{1-\cos{(\mu p_+)}}{\mu_+^2}+\frac{1-\cos{(\mu p_-)}}{\mu_-^2}+\\&-A^2(\tau)\hslash^2\bigg(\frac{\partial^2}{\partial p_+^2}+\frac{\partial^2}{\partial p_-^2}\bigg)\bigg]\chi(p_\pm,\tau)=i\hslash\frac{\partial \chi(p_\pm,\tau)}{\partial \tau}\,,
\end{align}
i.e. a time-dependent quantum pendulum. However, the hypothesis $\beta_+,\beta_-\ll1$ makes it possible to consider the regime of small oscillations in the pendulum dynamics, i.e. the aforementioned harmonic oscillator. 

\section{Quantum behaviour of the anisotropies for the Bianchi I model}\label{BI}
In this section we first analyze the anisotropies quantum behaviour by neglecting the quadratic potential, i.e. by restricting the cosmology to the simpler Bianchi I model. In this case, the Schr\"{o}dinger equation in the coordinate representation reduces to\footnote{In the following we have set $\hslash=1$ to simplify the numerical computations.}
\begin{equation}
\label{SBI}
-\bigg(\frac{\partial^2}{\partial \beta_+^2}+\frac{\partial^2}{\partial \beta_-^2}\bigg)\chi(\beta_\pm,\tau)=i\frac{\partial \chi(\beta_\pm,\tau)}{\partial \tau}
\end{equation}
whose solution can be written as a linear combination of
\begin{equation}
\varphi(\beta_\pm,\tau)=e^{ik_+\beta_+}e^{ik_-\beta_-}e^{-i(k_+^2+k_-^2)\tau}
\end{equation}
Then, we construct a Gaussian wave packet ($\mathcal{N}$ is the normalization constant) 
\begin{align}
\label{chi}
\chi(\beta_\pm,\tau)=\frac{1}{\mathcal{N}}\int\int_{-\infty}^{+\infty}&e^{-\frac{(k_+-\bar{k}_+)^2}{2\sigma_+^2}}e^{-\frac{(k_--\bar{k}_-)^2}{2\sigma_-^2}}\times\\&\times e^{ik_+\beta_+}e^{ik_-\beta_-}e^{-i(k_+^2+k_-^2)\tau}dk_+\,dk_-
\nonumber
\end{align}
and we study the evolution of the probability density distribution $|\chi|^2=\chi(\beta_\pm,\tau)\chi^*(\beta_\pm,\tau)$. Actually, the position of the peak and the width of the probability function give the information about the quantum behaviour of the anisotropies, i.e. their mean values and standard deviation. 

We notice that the presence of the Bounce enters in the definition of the time variable, since the relation between the synchronous time $t$ and the time variable $\tau$ is fixed by the choice of the lapse function $N=2A^{3/2}/\sqrt{2}$ and hence it depends on how the Universe volume evolves with time. Thus, polymer effects on the anisotropies can appear when returning to the synchronous time picture. In particular, in the bouncing picture we have (see \eqref{sol}) 
\begin{equation}
\label{cosh}
V^{pol}(\tau)=A(\tau)^{3/2}\,,
\end{equation}
whereas in the absence of the semiclassical polymer scenario we get the two singular trajectories (depending on the sign of the initial condition on $p_A$)
\begin{equation}
\label{exp}
V^\pm(\tau)=A^{\pm}(\tau)^{3/2}=(e^{\pm \frac{|p_A(0)|\tau}{\sqrt{288}\pi^2}})^{3/2}
\end{equation}
that can be obtained by taking the limit of the polymer solution \eqref{cosh} for $\mu\rightarrow0$. Therefore, by using the relation \eqref{tau} we have 
\begin{equation}
\label{tpol}
t=\int \frac{2V^{pol}(\tau)}{\sqrt{2}}\,d\tau
\end{equation}
 in the polymer case and
\begin{equation}
\label{tsing}
t=\int \frac{2V^\pm(\tau)}{\sqrt{2}}\,d\tau
\end{equation}
for the two singular branches. Thus, after computing the integrals, the obtained functions can be inverted to find the relation between $\tau$ and $t$ in the two cases. 
\begin{figure}[ht]
\centering
\includegraphics[width=0.85\linewidth]{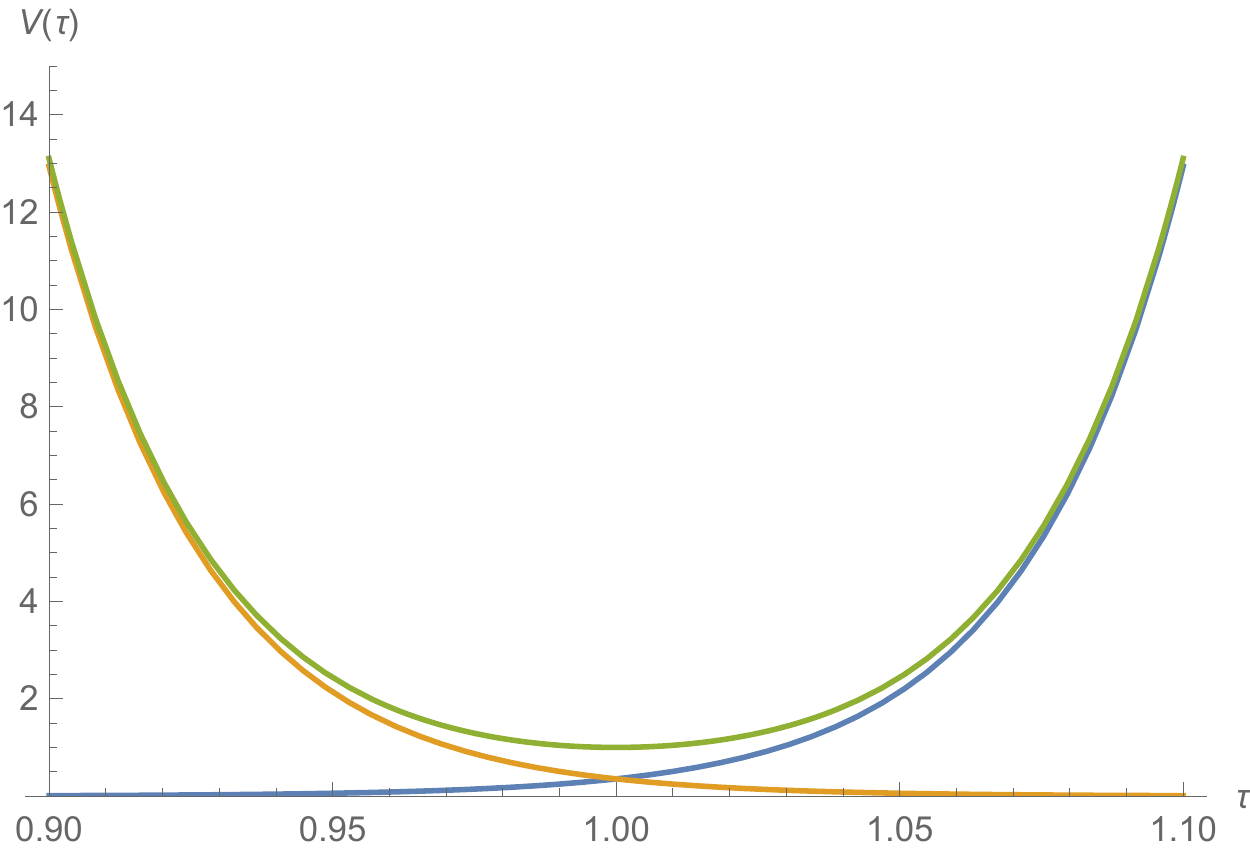}
\caption{Plot of the bouncing trajectory for the Universe volume in function of $\tau$, together with the trajectories for the classical limits $\tau\rightarrow\pm\infty$.}
\label{vtau}
\end{figure}

\begin{figure}[ht]
    \centering
\includegraphics[width=0.9\linewidth]{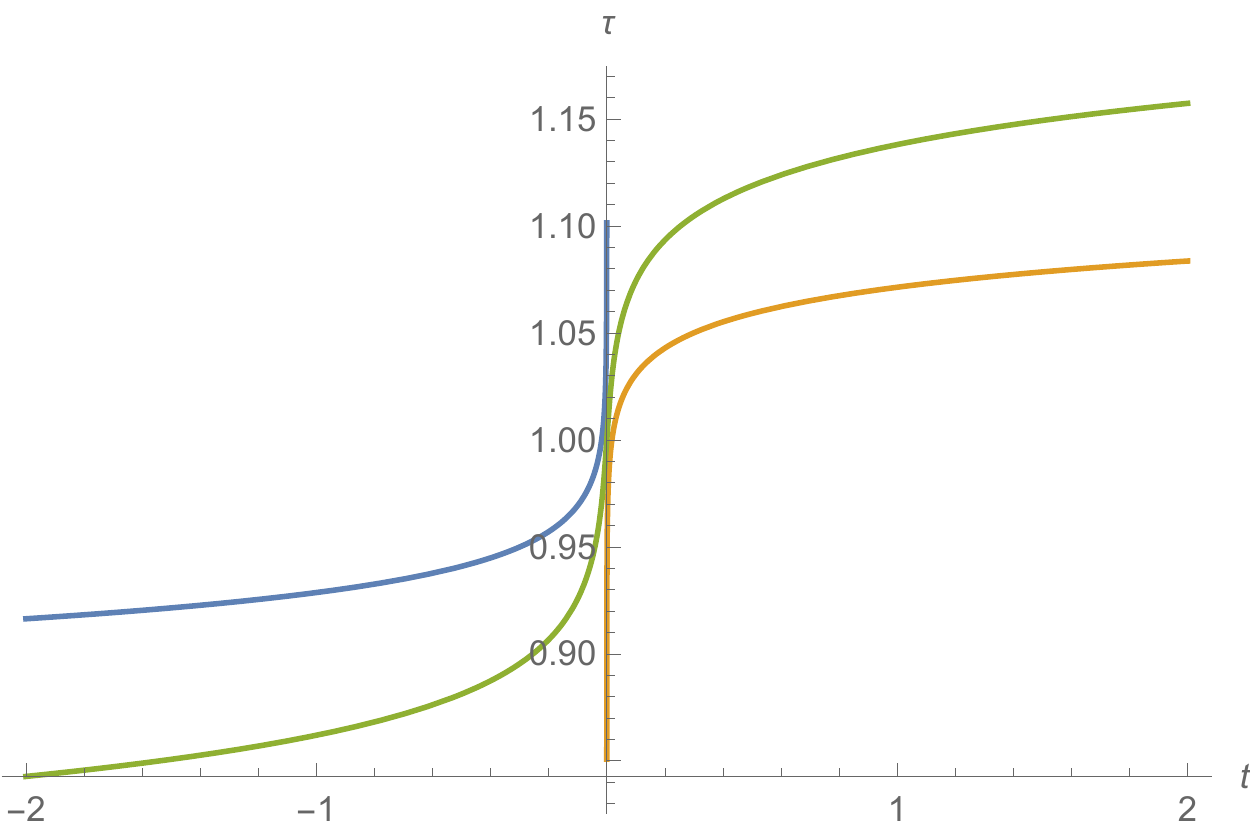}
    \caption{Plot of $\tau(t)$ for the three relevant cases: $\tau^{pol}(t)$ (green line) for the bouncing trajectory, $\tau^+(t)$ (orange line) for the expanding branch and $\tau^-(t)$ (blue line) for the collapsing one.}
    \label{Tau}
\end{figure}

In order to provide a generic bouncing scenario, here the initial conditions are such that the Bounce occurs at $\phi=1$ (see Fig. \ref{vtau}). Actually, the trajectory for the Universe volume is formally equivalent to the translation of \eqref{cosh} for $\tau\rightarrow\tau-1$. We note that the exponential behaviour of the classical limit is recovered for large times. Then, by means of numerical and analytical methods respectively we have found the inverse  functions $\tau^{pol}(t)$ and $\tau^\pm(t)$. In particular, from Fig. \ref{Tau} we point out that the divergence of $\tau^\pm(t)$ in correspondence of the singularity $t=0$ is regularized in the bouncing trajectory $\tau^{pol}(t)$. Moreover, $\tau^+(t)$ is not simply the reflection of $\tau^-(t)$ with respect to the bisector $\tau=t$ since the Bounce does not occur at $\phi=0$. 

\begin{figure}[ht]
\includegraphics[width=0.85\linewidth]{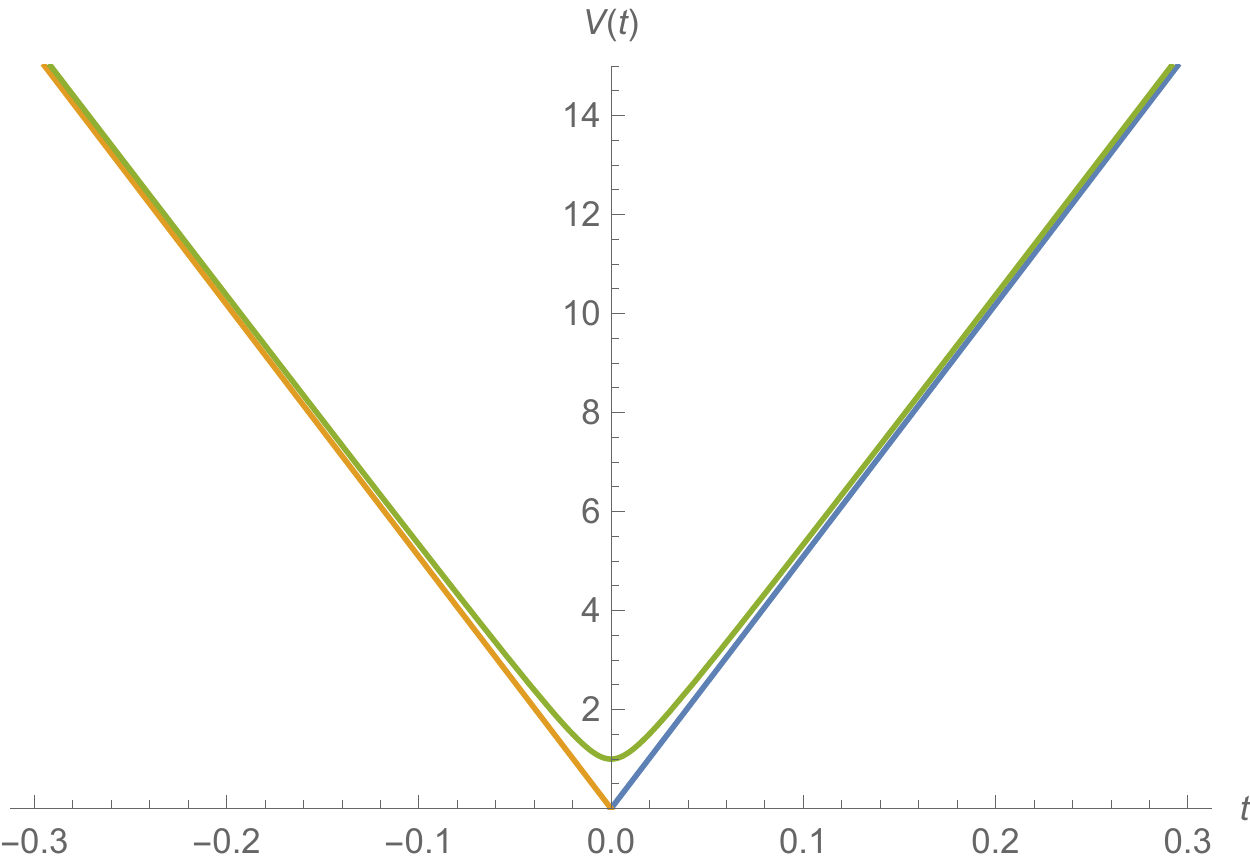}
\caption{Plot of the bouncing trajectory for the Universe volume in function of $t$, together with the trajectories for the classical limits $t\rightarrow\pm\infty$.}
\label{vt}
\end{figure}

For the sake of completeness, we also report the Universe volume trajectory in the synchronous time obtained by using the numerical solution $\tau^{pol}(t)$ (see Fig. \ref{vt}). We notice that the Bounce occurs for $t=0$ and that the linear behaviour of the classical solution is recovered for large times.

Now, by using $\tau^{pol}(t)$ and $\tau^{\pm}(t)$ in \eqref{chi} we have studied the two different evolution of the probability density function $|\chi(\beta_\pm,t|^2$ in the synchronous time, thus recovering the quantum information on the anisotropies in the bouncing picture and in the singular one respectively. Actually, it is well-known that the Schr\"{o}dinger solutions are characterized by a spreading behaviour and, restricting to the one-dimensional picture, it can be easily demonstrated that the Gaussian standard deviation grows linearly with time. Since \eqref{SBI} is separable, the total Universe wave function \eqref{chi} is factorized as $\chi(\beta_\pm,t)=\chi^+(\beta_+,t)\chi^-(\beta_-,t)$ in which
\begin{equation}
\chi^+(\beta_+,t)=\frac{1}{\mathcal{N}_+}\int_{-\infty}^{+\infty}e^{-\frac{(k_+-\bar{k}_+)^2}{2\sigma_+^2}}e^{ik_+\beta_+}e^{-ik_+^2t}dk_+
\end{equation}
and
\begin{equation}
\chi^-(\beta_-,t)=\frac{1}{\mathcal{N}_-}\int_{-\infty}^{+\infty}e^{-\frac{(k_--\bar{k}_-)^2}{2\sigma_-^2}}e^{ik_-\beta_-}e^{-ik_-^2t}dk_-\,.
\end{equation}
Hence, we can restrict the analysis to a single anisotropic degree of freedom and same considerations will be valid for the other one. Thus, if we consider the anisotropy $\beta_+$ we have that $\sigma_+(t)\sim\tau(t)$ and therefore in the singular picture the standard deviation of the Universe wave function $\chi^+(\beta_+,t)$ diverges towards the singularity $t\rightarrow0$ (that corresponds to the initial Big Bang in the singular expanding branch and to the future Big Crunch in the singular collapsing one). On the other hand, in the bouncing picture the divergence is regularized at the Bounce $t=0$ in which the two branches are symmetrically reconnected. However, the Universe wave packet is still affected by the spreading behaviour and the anisotropic standard deviation is not confined but it grows along the dynamics, even if the Universe initial state is sufficiently localized before the bouncing point.
\begin{figure}[h!]
    \centering
\includegraphics[width=1\linewidth]{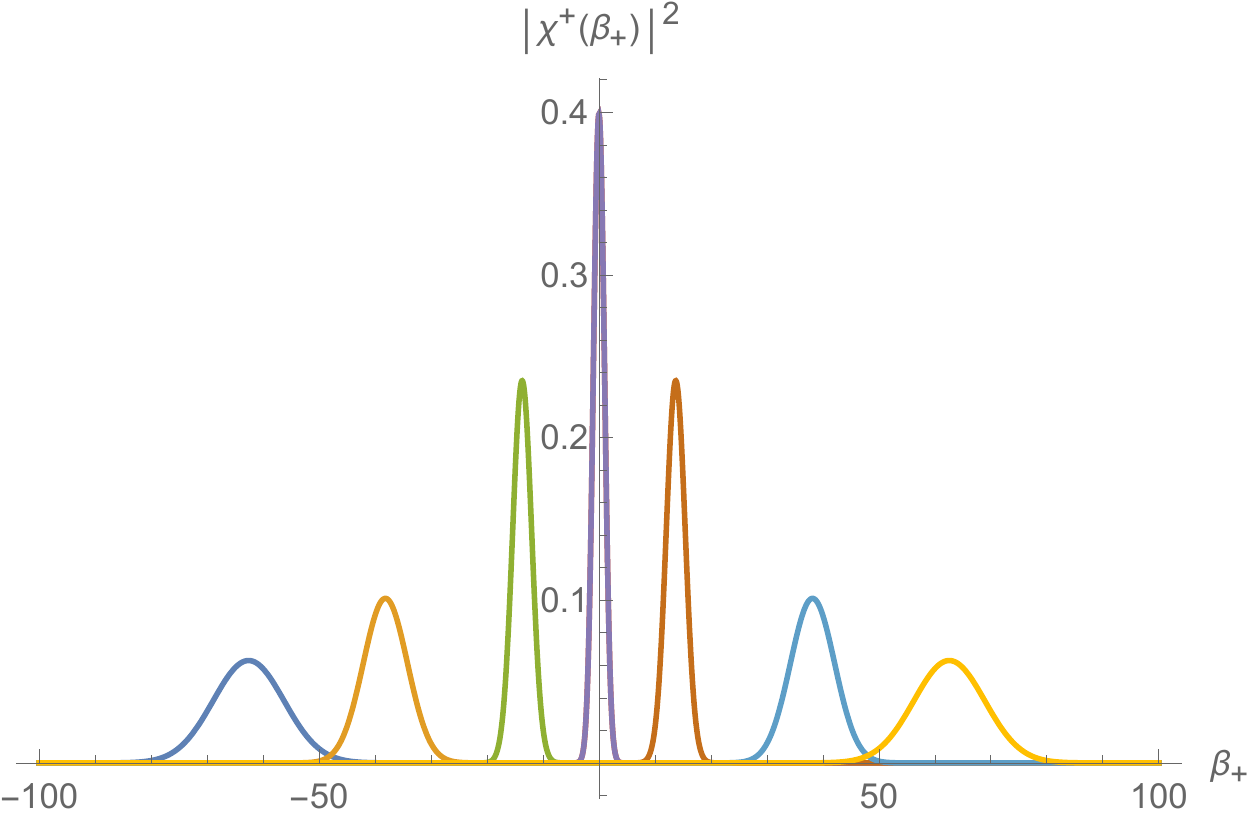}
    \caption{Left side: Evolution of $|\chi^+(\beta_+,\tau^+(t))|^2$ in function of the anisotropy $\beta_+$ from the initial singularity (for $t=10^{-180},10^{-120},10^{-60}$). Right side: Evolution of $|\chi^+(\beta_+,\tau^-(t))|^2$ in function of the anisotropy $\beta_+$ towards the future singularity (for the corresponding symmetric values of time).}
    \label{spack}
\end{figure}
\begin{figure}[ht]
\includegraphics[width=1\linewidth]{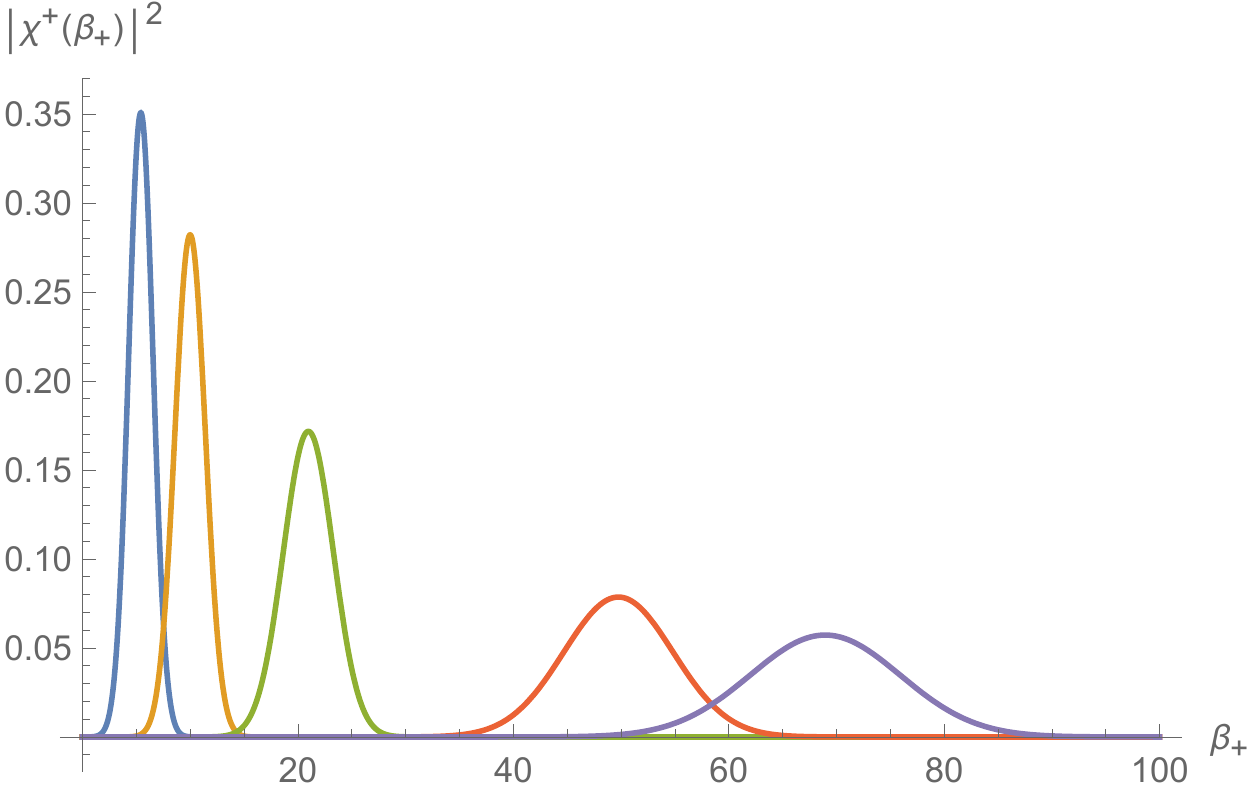}
\caption{Evolution of $|\chi^+(\beta_+,\tau^{pol}(t))|^2$ in function of the anisotropy $\beta_+$ in the bouncing scenario for $t=-10^5,0,10^{15},10^{60},10^{90}$).}
\label{both}
\end{figure}
\begin{figure*}
\begin{minipage}[ht]{7cm}	\includegraphics[width=1.1\linewidth]{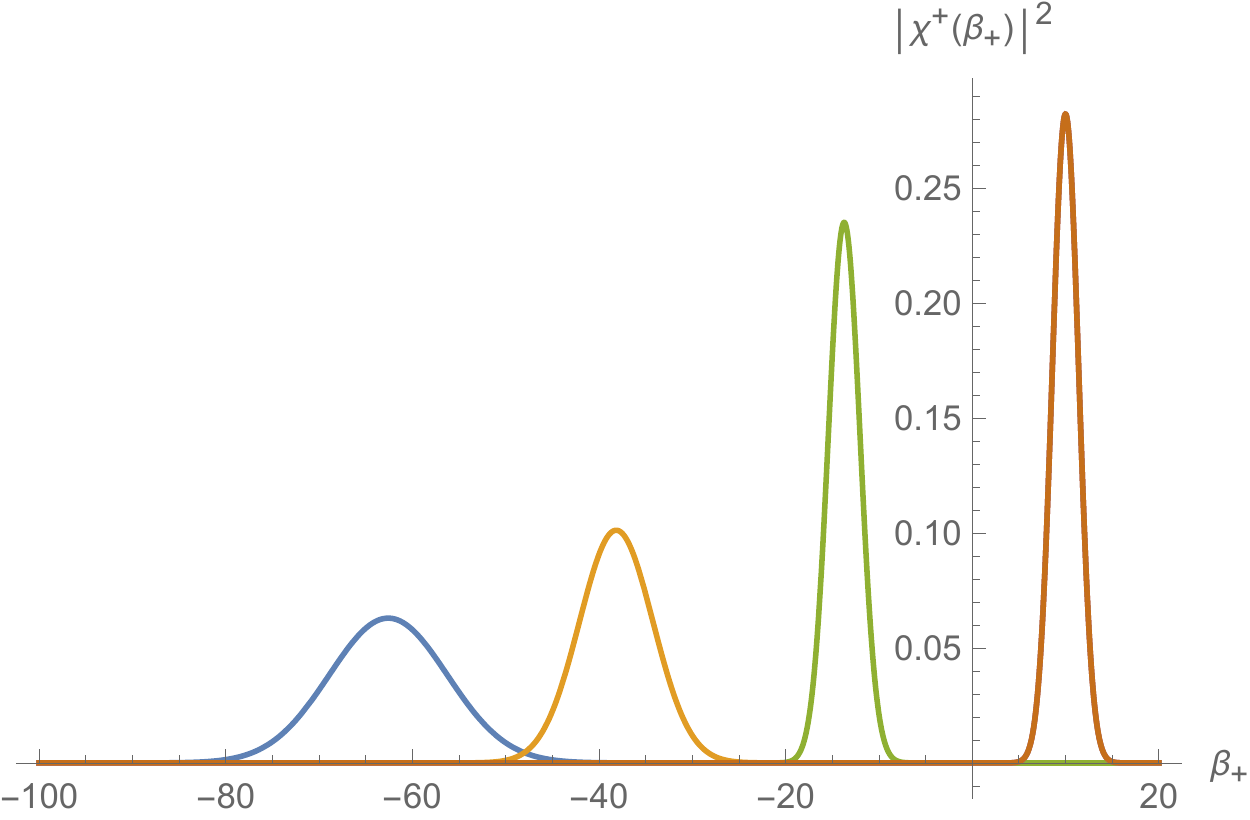}
\end{minipage}\qquad\qquad\qquad\qquad
\begin{minipage}[ht]{7cm}
\includegraphics[width=1.1\linewidth]{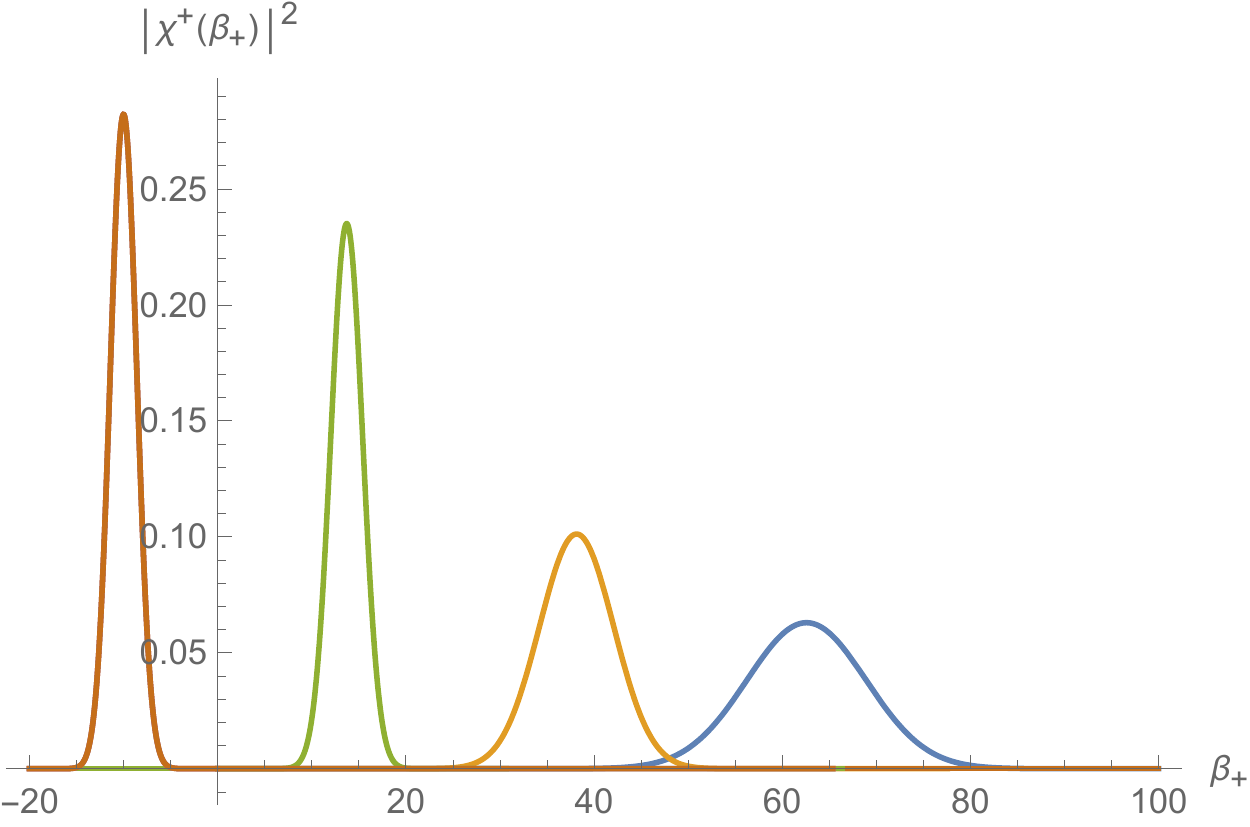}
\end{minipage}
\caption{Evolution of $|\chi^+(\beta_+,\tau^+(t))|^2$ in function of the anisotropy $\beta_+$ along the expanding branch for $t=10^{-180},10^{-120},10^{-60}$ (left/right side in the left/right figure) compared with that of $|\chi^+(\beta_+,\tau^{pol}(t))|^2$ in function of the anisotropy $\beta_+$ along the bouncing trajectory for $t=10^{-180},10^{-120},10^{-60}$ (right/left side in the left/right figure). The evolution of the Universe wave packet in the bouncing scenario is not appreciable at these time scales. For the Gaussian coefficients we have set $\sigma_+=1/\sqrt{2}$, $\bar{k}_+=+5$ in the left figure and $\bar{k}_+=-5$ in the right figure.}
\label{compare}
\end{figure*}

In Fig. \ref{spack} we can see the behaviour of the Universe wave packet along the expanding branch (left side) and collapsing one (right side) that shows the spreading behaviour towards the singularities. From the position of the peak we can infer that the more the wave packet spreads the higher the anisotropy mean value is (in absolute value). 

When considering the bouncing evolution for the semiclassical sector, we have that the Universe wave packet spreads slowly (see Fig. \ref{compare}) and therefore the anisotropic standard deviation can remain confined for a finite time interval if the initial Universe wave packet is sufficiently localized. However, even if the quantum behaviour is regularized at the Bounce the anisotropic standard deviation grows indefinitely for $t\rightarrow+\infty$ (see Fig. \ref{both}). In particular, states with higher anisotropic standard deviation are associated to higher (absolute) mean values for the anisotropy. Finally, we repeat the analysis above considering the anisotropies as discrete. Accordingly, the Sch\"{o}dinger equation in the semiclassical polymer framework writes as
\begin{equation}
    \bigg[\frac{\sin^2({\mu_+p_+})}{\mu_+^2}+\frac{\sin^2({\mu_-p_-})}{\mu_-^2}\bigg]\chi_{pol}(p_\pm,\tau)=i\frac{\partial \chi_{pol}(p_\pm,\tau)}{\partial \tau}\,,
\end{equation}
in which we have recovered the momentum representation.
Thus, after a Fourier transformation the solution can be written as a factorized Gaussian superposition of plane waves
\begin{align}
\nonumber
\chi_{pol}(\beta_\pm,\tau)&=\chi_{pol}^+(\beta_+,t)\chi_{pol}^-(\beta_-,t)=\\&\nonumber=\frac{1}{\mathcal{N}}\int\int_{-\infty}^{+\infty}e^{-\frac{(k_+-\bar{k}_+)^2}{2\sigma_+^2}}e^{-\frac{(k_--\bar{k}_-)^2}{2\sigma_-^2}}\times\\&\times e^{ik_+^\mu\beta_+}e^{ik_-^\mu\beta_-}e^{-i\left[(k_+^\mu)^2+(k_-^\mu)^2\right]\tau}dk_+\,dk_-
\end{align}
in which $\mathcal{N}$ sets the norm and $k^\mu_\pm=\sin(\mu_\pm k_\pm)/\mu_\pm$. Now we can study the evolution of the probability density $|\chi_{pol}^+(\beta_+,t)|^2$ as before, and analyze the different behaviour in the singular or bouncing background picture by respectively considering $\tau^\pm(t)$ or $\tau^{pol}(t)$. 
\begin{figure}[b]
    \centering
\includegraphics[width=1\linewidth]{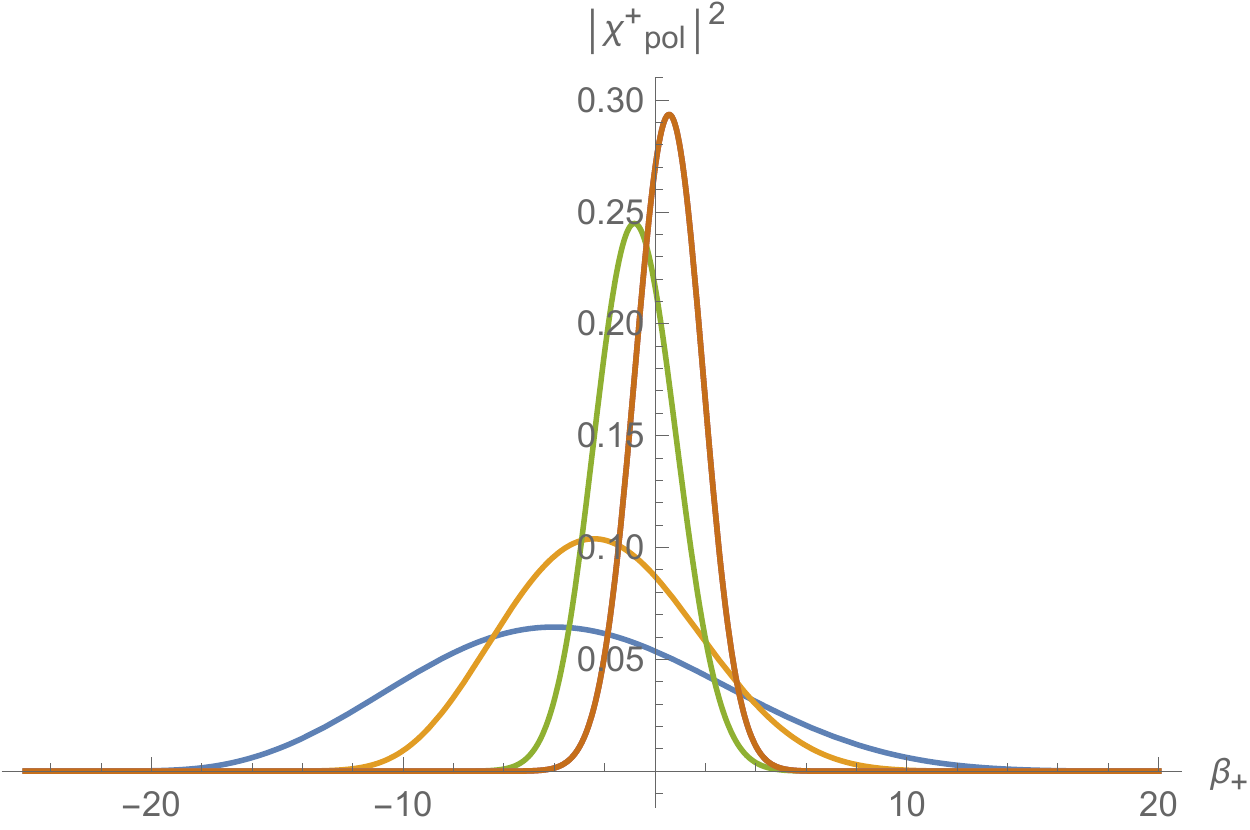}
    \caption{Evolution of $|\chi_{pol}^+(\beta_+,\tau^+(t))|^2$ in function of the anisotropy $\beta_+$ along the expanding branch (left side of the figure) for $t=10^{-180},10^{-120},10^{-60}$ compared with that of $|\chi_{pol}^+(\beta_+,\tau^{pol}(t))|^2$ in function of the anisotropy $\beta_+$ along the bouncing trajectory (red lines) for $t=10^{-180},10^{-120},10^{-60}$ (we set $\sigma_+=1/\sqrt{2}$, $\bar{k}_+=-5$).}
    \label{polani1}
\end{figure}

\begin{figure}[h]
    \centering
    \includegraphics[width=1\linewidth]{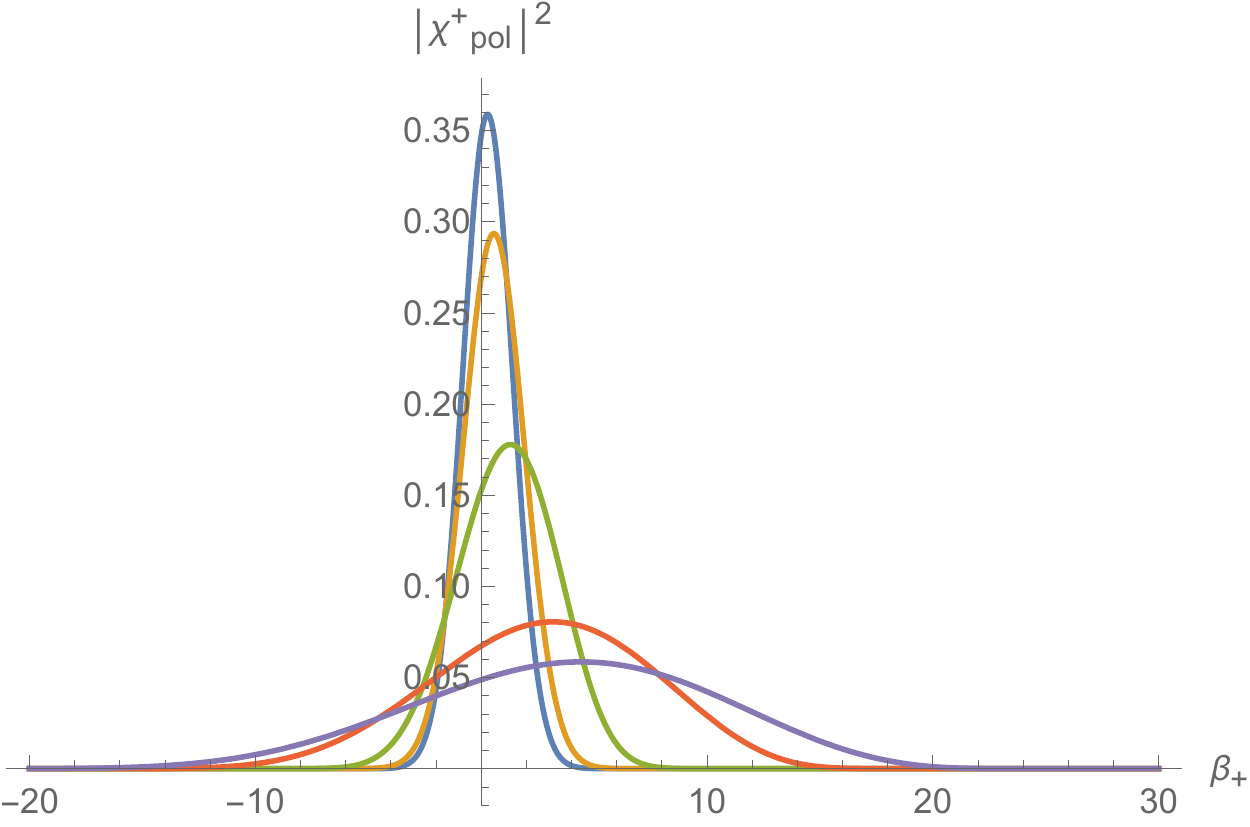}
    \caption{Evolution of $|\chi_{pol}^+(\beta_+,\tau^{pol}(t))|^2$ in function of the anisotropy $\beta_+$ in the bouncing scenario for $t=-10^5,0,10^{15},10^{60},10^{90}$).}
    \label{polani2}
\end{figure}

In particular, in Fig. \ref{polani1} we can see the behaviour of the Universe wave packet along the expanding branch (left side) and the bouncing one (red lines), whose evolution in the latter case is not appreciable at these time scales. More specifically, in Fig. \ref{polani2} it is highlighted the slower, but still present, spreading behaviour of the bouncing case too. Thus, the unlimited growth of the anisotropies standard deviation is not prevented even when they are represented in the polymer formalism. 

\section{Quantum behaviour of the anisotropies for the Bianchi IX model}\label{BIX}
\begin{figure*}
\begin{minipage}[ht]{7cm}	\includegraphics[width=1\linewidth]{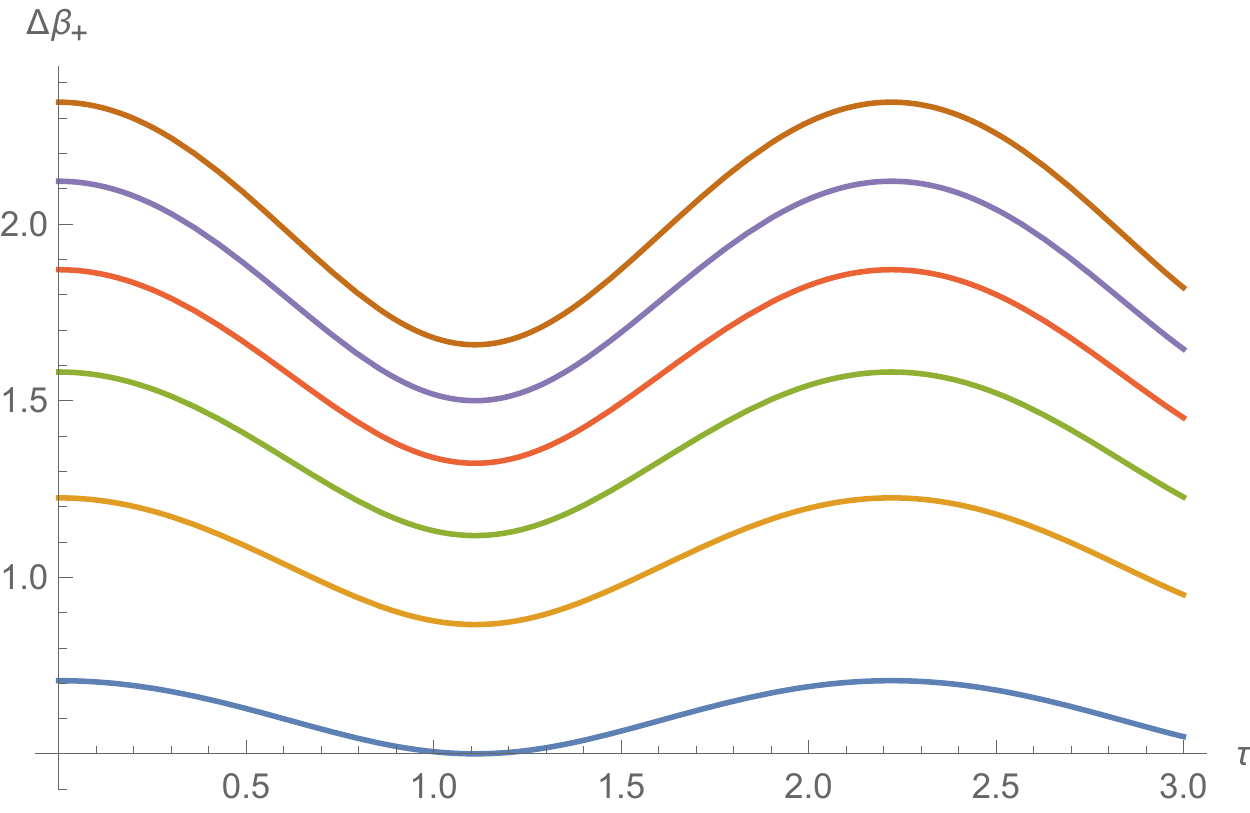}
\end{minipage}\qquad\qquad\qquad\qquad
\begin{minipage}[ht]{7cm}
\includegraphics[width=1\linewidth]{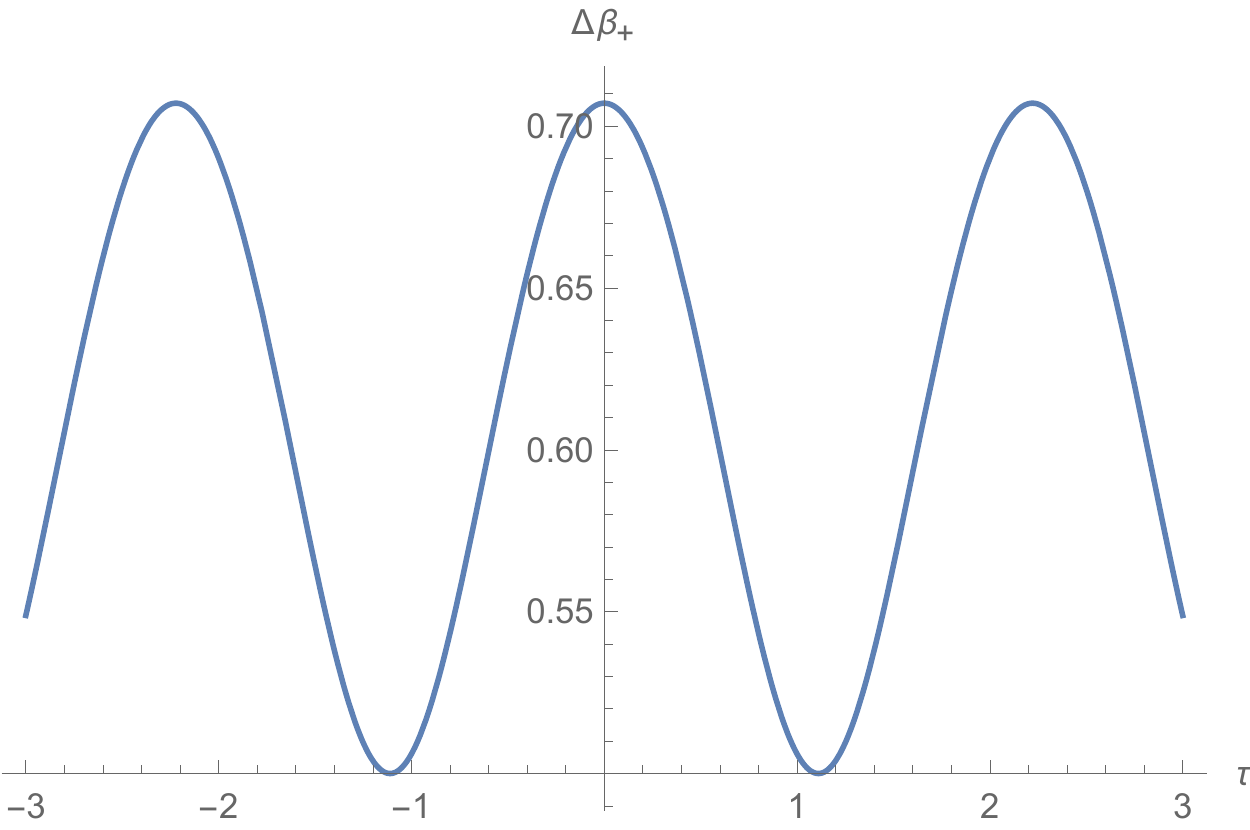}
\end{minipage}
\caption{Left figure: Behaviour of $\Delta(\beta_+)$ in function of $\tau$ for $n=0,1,2,3,4,5$ in the semiclassical bouncing picture. Right figure: Behaviour of $\Delta(\beta_+)$ in function of $\tau$ for $n=0$ in the bouncing picture.}
\label{osc}
\end{figure*}
In this section we analyze the anisotropies quantum behaviour in the full model, i.e. by including the harmonic potential $A^2(\tau)(\beta_+^2+\beta_-^2)$. The analytical method to solve the time-dependent quantum oscillator (see \eqref{S}) is widely described in \cite{Pedrosa}. In particular, by means of the exact invariant method it can be demonstrated that the normalized eigenfunctions of the Hamiltonian can be written as
\begin{equation}
    \chi_n(\beta_+,\beta_-,\tau)=\chi_n^+(\beta_+,\tau)\chi_n^-(\beta_-,\tau)
\end{equation}
with
\begin{equation}
\chi_n^\pm(\beta_\pm,\tau)=\frac{e^{i\gamma_n(\tau)}}{\sqrt{\sqrt{\pi}n!2^n\rho}}\mathcal{H}\bigg(\frac{\beta_\pm}{\rho}\bigg)e^{\frac{i}{4}(\frac{\dot{\rho}}{\rho}+\frac{2i}{\rho^2})\beta_\pm^2}
\end{equation}
in which $\mathcal{H}$ are the usual Hermite polynomials, the $\gamma$ function is defined as
\begin{equation}
\label{gamma}
\gamma_n(\tau)=-\bigg(n+\frac{1}{2}\bigg)\int_0^\tau\frac{2}{\rho^2}d\tau'
\end{equation}
and $\rho$ solves the differential equation
\begin{equation}
\label{rho}
    \ddot{\rho}+4A(\tau)^2\rho=0\,.
\end{equation}
Actually, in our model the role of the time-dependent frequency is played by the solution for the Universe area $A(\tau)$. Thus, the semiclassical behaviour of the model  directly enters in the quantum dynamics thanks to the presence of the Bianchi IX potential. 
\begin{figure}[ht]
    \centering
\includegraphics[width=1\linewidth]{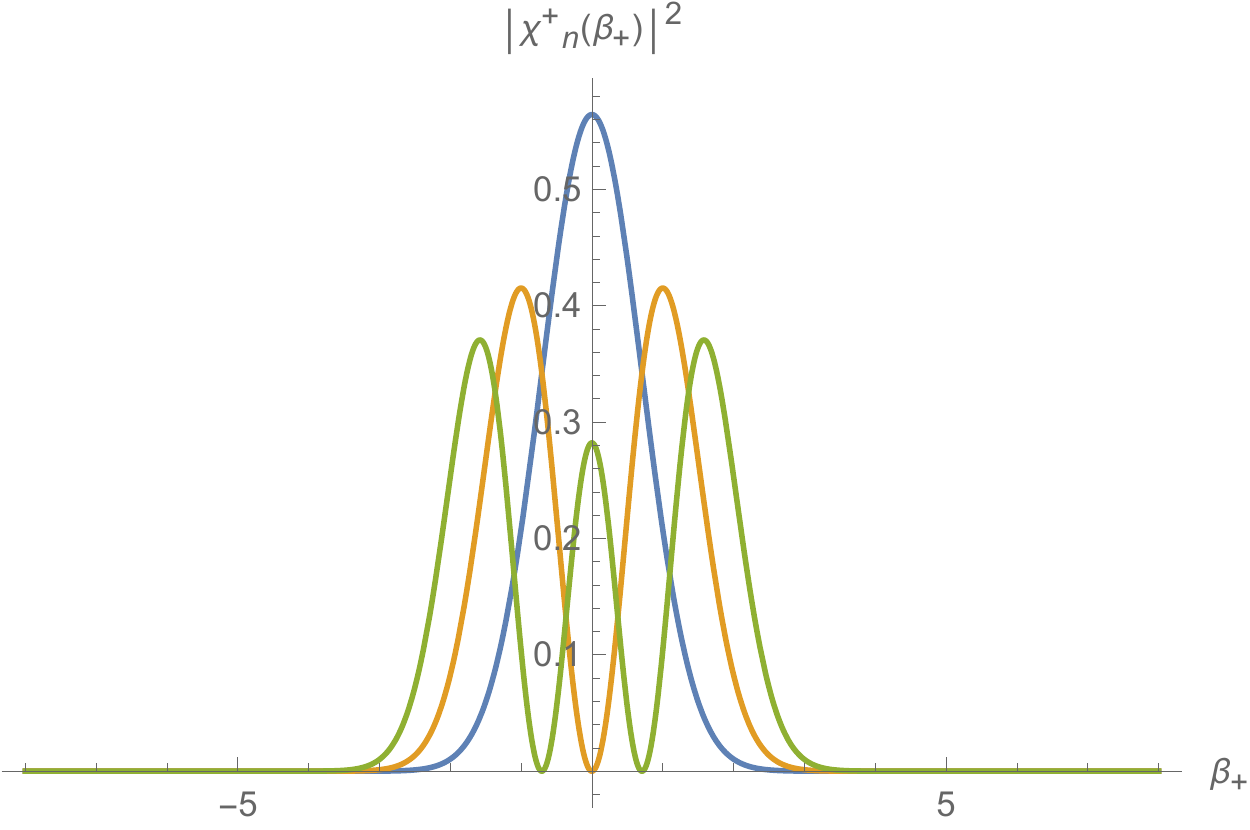}
    \caption{Plot of $|\chi(\beta_+,0)|^2$ for $n=0$ (blue line), $n=1$ (orange line) and $n=2$ (green line).}
    \label{Psi}
\end{figure}

By solving \eqref{rho} (in which for $A(\tau)$ we have used the expression in \eqref{sol} with $\mu=1/3$) we get
\begin{equation}
\rho(\tau)=c_1\mathcal{M}_C\bigg(-\frac{1}{288},\frac{1}{576},24i\tau\bigg)-c_2\mathcal{M}_S\bigg(-\frac{1}{288},\frac{1}{576},24i\tau\bigg)\,,
\end{equation}
where $c_1,c_2$ are constants and $\mathcal{M}_C$ and $\mathcal{M}_S$ are the Mathieu functions, i.e. elliptic cosine and the elliptic sine respectively. Then, we have constructed a basis of real solutions and performed an expansion of the elliptic functions up to the first order, i.e.
\begin{align}
\label{MC}
    &\mathcal{M}_C\bigg(-\frac{1}{288},\frac{1}{576},24i\tau\bigg)\sim\cos\bigg(\sqrt{-\frac{1}{288}}24i\tau\bigg)\,,\\
    \label{MS}
    &\mathcal{M}_S\bigg(-\frac{1}{288},\frac{1}{576},24i\tau\bigg)\sim\sin\bigg(\sqrt{-\frac{1}{288}}24i\tau\bigg)\,.
\end{align}
This approximation is justified since the second argument of $\mathcal{M}_{C,S}$ tends to zero in the continuum limit, i.e. for $\mu\ll1$. Thanks to the expansions \eqref{MC} and \eqref{MS} we are able to analytically compute the integral in \eqref{gamma}. Thus, we have an explicit expression for the eigenstates $\chi_n(\beta_+,\beta_-,\tau)$ through which computing the mean value and the standard deviation of the anisotropies $\beta_\pm$. 

In Fig. \ref{Psi} we show the probability density distribution $|\chi_n^+(\beta_+,0)|^2$ for the obtained eigenfunctions in the anisotropy $\beta_+$ with $n=0,1,2$. We also remark that the mean value of the coordinate (here the anisotropy) is zero on the harmonic oscillator eigenstates, hence to obtain the behaviour of the standard deviation we have to compute the mean value of $\beta_+^2$, i.e.
\begin{equation}
<\beta_+^2>=\int_{-\infty}^{+\infty}(\chi^+_n(\beta_+,\tau))^*\beta_+^2\chi^+_n(\beta_+,\tau)\,d\beta_+\,,
\end{equation}
and then $\Delta\beta_+=\sqrt{<\beta_+^2>}$. Obviously, all the following analysis is analogous if performed for $\beta_-$. In particular, we have derived the standard deviation $\Delta\beta_+$ for the first six energy levels, including the lower one. 

In Fig. \ref{osc} it is highlighted the confined character of $\Delta\beta_+$ in function of the time variable $\tau$. In particular, $\Delta\beta_+$ oscillates around a constant values that grows with $n$ but remains one order of magnitude smaller than $\sigma_+$ in the case of the two-dimensional free particle, i.e. Bianchi I (see Fig. \ref{compare} in which both the singular and the bouncing pictures are reported). Thus, the presence of the harmonic potential maintains the anisotropies small, differently from the previous case (see Sec. \ref{BI}). 

\begin{figure}[ht]
    \centering
\includegraphics[width=1\linewidth]{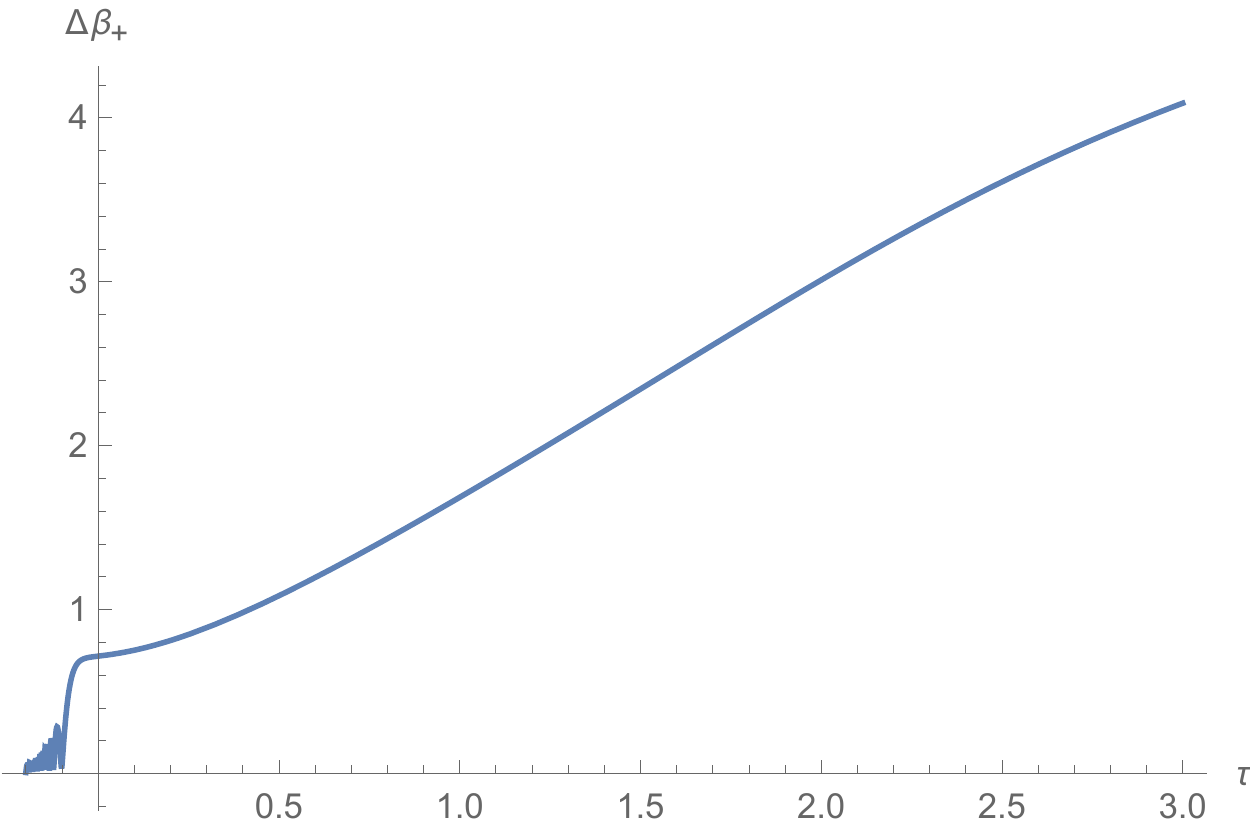}
    \caption{Behaviour of $\Delta(\beta_+)$ in function of $\tau$ for $n=0$ along the collapsing (singular) branch.}
    \label{sing}
\end{figure}

The corresponding behaviour of $\Delta\beta_+$ in the absence of the regularized semiclassical bouncing evolution is represented in Fig. {sing}. In this case, we have considered $A^-(\tau)$ (see \eqref{exp}) as the time-dependent frequency of the harmonic oscillator in \eqref{S}, i.e. the collapsing singular solution obtained from the considered bouncing solution $A(\tau)$ in the limit $\mu\rightarrow0$. Then, by following the same procedure explained above we have found the complete set of eigenstates of the corresponding Hamiltonian. In particular, \eqref{rho} turn to be a Bessel-type one and therefore we can write
\begin{equation}
    \rho(\tau)=c_1\mathcal{B}_J(0,A^-(\tau)/6\pi)+c_22\mathcal{B}_Y(0,A^-(\tau)/6\pi)\,,
\end{equation}
in which $\mathcal{B}_J$ and $\mathcal{B}_Y$ are the Bessel functions of the first and second kind respectively. Therefore, $\Delta\beta_+$ loses its oscillatory and confined character and grows monotonically towards the singularity (see Fig. \ref{sing}).

Thus, we can conclude that the only presence of the harmonic potential is not sufficient to confine the trajectory of the anisotropies but it is necessary to introduce the regularizing polymer effects and remove the divergences in the semiclassical background.

\section{Concluding remarks}\label{C}

In this work we consider a new theoretical paradigm to investigate the decoupling of a quantum gravity system into a quasi-classical background and a small quantum subsystem. In particular, we generalize the original idea in \cite{Vilenkin89} in the case of a polymer quantum formulation of the Minisuperspace \cite{Corichi,Barca21}. The main goal of the present analysis is the quantum characterization of the Universe anisotropic degrees of freedom across a Big Bounce configuration \cite{Ashtekar11,Mantero18}. 

The first part of the manuscript is dedicated to the challenging question of reformulating the Vilenkin idea in the momentum representation, i.e. the only viable to semiclassically implement the polymer formulation. In fact, the Hamiltonian kinetic term is always quadratic in the momenta and hence in the standard Vilenkin approach it transforms into a second derivative in the coordinate representation. On the other hand, the potential term has a generic form and so in the momentum representation we have to deal with non-local differential operators. Nonetheless, we manage this difficulty by requiring that all the functions of the momenta be series expandable.

In the second part of this study we apply the derived theoretical paradigm to the description of the quantum behaviour of small Universe anisotropies during a semiclassical bouncing picture. 
In particular, we first consider a Bianchi I model (to be thought as the limit of any Bianchi Universe when the spatial curvature is neglected \cite{KL63,Imponente01}), whose background corresponds to a flat isotropic Universe. Then, we consider the evolution of a Bianchi IX model in the limit of small anisotropies, whose quantum potential contains the positive spatial curvature contribution.

In the first case, we see that thanks to the presence of the Bounce the anisotropies quantum standard deviation grows slowly without diverging at $t=0$ (differently from what happens along the singular trajectories) but it still monotonically increases. So, the concept of a quasi-isotropic Universe (as well as the Vilenkin approximation scheme) is essentially lost since the anisotropies are not under control, even when represented on the polymer lattice. 

In the second case, the resulting Schr\"{o}dinger equation for the anisotropic variables in the polymer formalism would correspond to a time-dependent quantum pendulum, reduced to a simpler and treatable time-dependent harmonic oscillator by virtue of the smallness of the anisotropic phase space, i.e. the fundamental assumption of the present Vilenkin analysis. The important result is that the anisotropies standard deviation of the Hamiltonian eigenstates is no longer monotonically increasing across the Bounce, but it oscillates with a constant amplitude. The main difference between the two cases is that the polymer bouncing effect alone is not able to maintain the smallness of the anisotropies (both of their mean value and standard deviation), whereas the quantum emergence of a small positive curvature confines the anisotropies standard deviation if the quasi-classical background is regularized at the singularity.

Thus, by limiting the anisotropies to a small quantum effect we have demonstrated that the bouncing evolution preserves the quasi-isotropic nature of the classical background, thanks to the presence of a spatial curvature term that induces a localizing potential. We remark that the anisotropic mean values on the Hamiltonian eigenstates is zero, hence our analysis is consistent with the quasi-isotropic ansatz. By concluding, since the Bianchi IX model is the prototype for constructing the dynamics of a generic inhomogeneous cosmological model \cite{BKL82,Kirillov92,Montani95,Benini04} we can infer the general validity of the present analysis.
\acknowledgement{\textbf{Acknowledgements}}
We would like to thank Dr. Valerio
Cascioli, who first tried to address this problem during his PhD Thesis (see the unpublished manuscript arXiv:1903.09417 [gr-qc]).

	\nocite{*}
\bibliographystyle{plain}
\bibliography{bib}
\end{document}